\documentclass[preprint,showpacs,preprintnumbers,amsmath,amssymb]{revtex4}
\usepackage{etex}
\usepackage{graphicx}% Include figure files
\usepackage{dcolumn}% Align table columns on decimal point
\usepackage{bm}% bold math
\usepackage{amssymb,amsthm,amscd,amsbsy,array}
\usepackage{amsmath,dsfont}
\usepackage{soul} 
\usepackage{graphics,xcolor}

\renewcommand\thesection{\arabic{section}}
\numberwithin{equation}{section}

\newcommand{\half}{{\frac{1}{2}}}
\def\2{{\half}}
\newcommand{\const}{\mathop{\rm const}\nolimits}
\def\bA{{\bm{A}}}

\def\p{{\partial}}

\def\bbeta{{\bm{\beta}}}
\def\bgamma{{\bm{\gamma}}}
\def\bomega{\mbox{\boldmath$\omega$}}
\def\bp{{\bm{p}}}
\def\tbp{\tilde{\bm{p}}}
\def\tbx{\tilde{\bm{x}}}
\def\tV{\widetilde{V}}

\def\ba{{\bm{a}}}
\def\bc{{\bf c}}
\def\bu{{\bm{u}}}

\def\bnabla{\mbox{\boldmath$\nabla$}}
\def\bTheta{\mbox{\boldmath$\Theta$}}
\def\br{{\bm{r}}}

\def\bg{{\bm{g}}}

\def\bE{{\bm{E}}}
\def\bB{{\bm{B}}}
\def\bb{{\bm{b}}}
\def\bD{{\bm{D}}}
\def\bnabla{{\bm{\nabla}}}
\def\bq{{\bm{q}}}
\def\bp{{\bm{p}}}

\def\bs{{\bm{s}}}

\def\hbp{{\widehat{\bm{p}}}}
\def\bx{{\bm{x}}}
\def\bz{{\bm{z}}}
\def\by{{\bm{y}}}
\def\bZ{{\bm{Z}}}

\newcommand{\np}{\vert\bp\vert}

\def\beq{\begin{equation}}
\def\eeq{\end{equation}}
\def\beqa{\begin{eqnarray}}
\def\eeqa{\end{eqnarray}}

\def\barray{\left(\begin{array}}
\def\earray{\end{array}\right)}
\def\barraynb{\begin{array}}
\def\earraynb{\end{array}}

\def\SO{{\rm SO}}
\def\ort{{\mathfrak{o}}}
\def\sort{{\mathfrak{so}}}

\def\smallover#1/#2{\hbox{$\textstyle\frac{#1}{#2}$}} %

\newcommand{\cC}{\mathcal{C}}

\newcommand{\cL}{\mathcal{L}}

\newcommand{\bbR}{\mathbb{R}}

\newcommand{\fm}{{\mathfrak{{m}}}}

\newcommand{\cE}{{\mathcal{E}}}

\newcommand{\belle}{\boldsymbol{\ell}}
\newcommand{\se}{\mathfrak{e}}
\newcommand{\rE}{\mathrm{E}}
\newcommand{\SE}{\mathrm{SE}}

\usepackage{color}
\newcommand{\red}{\textcolor{red}}  
\newcommand{\blue}{\textcolor{blue}}
\newcommand{\green}{\textcolor{green}}

\begin{document}

\preprint{arXiv:1406.0718v4 [hep-th]
}

\title{Chiral fermions as classical massless spinning particles\\
}

\author{
C. Duval$^{1}$\footnote{mailto:duval@cpt.univ-mrs.fr},
P. A. Horv\'athy$^{2,3}$\footnote{mailto:horvathy@lmpt.univ-tours.fr}
}

\affiliation{
${}^1$Aix-Marseille Universit\'e, CNRS CPT UMR 7332, 13288 Marseille, France\\
Universit\'e de Toulon, CNRS CPT UMR 7332, 83957 La Garde, France
\\
${}^2$Laboratoire de Math\'ematiques et de Physique
Th\'eorique,
Universit\'e de Tours, (France)
\\
$^3$Institute of Modern Physics, Chinese Academy of Sciences, Lanzhou, (China) 
}

\date{today}

\begin{abstract}
Semiclassical chiral fermion models with Berry term are studied in a symplectic framework. In the free case, the system can be obtained from Souriau's model for a relativistic massless spinning particle  by ``enslaving'' the spin. The Berry term is identified with the classical spin two-form of the latter model. The Souriau model carries a natural Poincar\'e symmetry that we highlight, but spin enslavement breaks the boost symmetry. However the relation between the models  allows us to derive a Poincar\'e symmetry of  unconventional form  
for chiral fermions. Then we couple our system to an external electromagnetic field. 
For gyromagnetic ratio $g=0$ we get curious superluminal Hall-type motions; for $g=2$ and in a pure constant magnetic field  in particular, we find instead spiraling motions.
\end{abstract}


\maketitle

\tableofcontents

%%%%%%%%%%%%%%%%%%%%%%%%%%%%%%%%%%%%%%%%%%%%%%%%%%%%%%%%%%%%%%%%%%%%%%%%%%%%%%
%%%%%%%%%%%%%%%%%%%%%%%%%%%%%%%%%%%%%%%%%%%%%%%%%%%%%%%%%%%%%%%%%%%%%%%%%%%%%%
\section{Introduction}
%%%%%%%%%%%%%%%%%%%%%%%%%%%%%%%%%%%%%%%%%%%%%%%%%%%%%%%%%%%%%%%%%%%%%%%%%%%%%%
%%%%%%%%%%%%%%%%%%%%%%%%%%%%%%%%%%%%%%%%%%%%%%%%%%%%%%%%%%%%%%%%%%%%%%%%%%%%%%

Massless chiral fermions have attracted considerable recent interest  
\cite{SonYama1,Stephanov,hadrons,Dunne,SonYama2,ChenWang,Stone,Chen2013,ChenSon,Manuel,NewStone,WS,Karabali}.
Sophisticated quantum calculations are  greatly simplified by using (semi)classical models
which can be derived from the Dirac equation \cite{Stephanov}.
The model proposed in \cite{SonYama1,Stephanov,Stone}, for example, describes a spin-$1/2$ system with positive helicity and energy, by the phase-space action
\beq
S=\int\Big(\big(\bp+e\bA\big)\cdot\frac{d{\bx}}{dt}-\big(|\bp|+
e\phi
\big)
-\ba\cdot\frac{d{\bp}}{dt}
\Big)dt,
\label{chiract}
\eeq
which also involves
 the additional momentum-dependent vector potential $\ba(\bp)$ for the Berry monopole in $\bp$-space \cite{Niu},
\beq
\bnabla_{\bp}\times\ba=\bTheta\equiv
\frac{\hbp}{2|\bp|^2},
\label{Berryterm}
\eeq
 where $\hbp$ is the unit vector $\hbp=\displaystyle{\bp}/{|\bp|}$. Here $\bA(\bx,t)$ and $\phi(\bx,t)$ are ordinary vector and scalar potentials and $e$ is the electric charge.   
 \goodbreak

A remarkable feature of the system (\ref{chiract}) is its \emph{lack of manifest Lorentz symmetry} even  in the absence of an external gauge field \footnote{Alternative equations with Lorentz symmetry were proposed in \cite{ChenWang,ChenSon}, and the model was generalized to the non-Abelian context \cite{Stone,Chen2013}.}.

In this paper we first show that a \emph{free} chiral fermion model can be related 
 to Souriau's \emph{relativistic model of a massless spinning particle} \cite{SSD} by ``enslaving'' the spin, 
\beq
\bs=s\,{\hbp}, 
\eeq
 cf. (\ref{enslavement}),
 viewed as a sort of gauge fixing condition.  
The massless spinning model carries a natural Poincar\'e symmetry,  which we also generalize to finite transformations. This natural symmetry  \emph{is not inherited by the chiral model}, though~: spin enslaving  breaks the Poincar\'e symmetry to the so-called Aristotle group \cite{SSD} spanned by rotations and by space- and time- translations : \emph{the chiral system carries no natural boost symmetry}.

The subtle relationship between the two models allows, nevertheless, for a \emph{different},   \emph{twisted  Poincar\'e symmetry} for the chiral fermion (\ref{chiract}), obtained by
exporting the one carried by the massless spinning model. We stress that this twisted  Poincar\'e symmetry should be considered rather as a \emph{dynamical symmetry} in that its action on space-variables also involves the momentum. 

Then we study the coupling to an external electromagnetic field.
Applying first Souriau's version of minimal coupling 
 \cite{SSD, Souriau74} to the massless spinning model, we obtain a rather peculiar system, described in Section \ref{minimalSec}, which exhibits superluminal velocities
with a Hall-type behavior both for $3$-space and spin motion. 
 
We consider next a more general, non-minimal coupling scheme, which accommodates any gyromagnetic ratio, $g$, by  allowing the mass-square to depend on the coupling between the spin and the electromagnetic field \cite{Duval75,Souriau74}.
 The resulting, rather complicated system, presented in Section  \ref{anomSec}, combines the equations of motion of the previously studied minimal model ($g=0$), with new, Stern-Gerlach-type terms, which involve derivatives of the field, reminiscent of recent propositions \cite{ChenSon,Manuel}.
 
For the normal value $g=2$, which is consistent with the Dirac equation \cite{Duval75}, the anomalous terms are switched off,
leading to considerable simplification. In a uniform,  purely magnetic field we find, for example, spiraling motions. Spin  enslavement, although not mandatory, is possible in this case.

\goodbreak

The same procedure applied to chiral fermions allows to recover the results in \cite{Stephanov,Stone}. But the  chiral and the massless spinning  systems behave differently because of the extra structure of the latter, which remains hidden in the free case, but which comes to light under coupling to an external field: the chiral model has a $6$-dimensional phase space, while the massless spinning particle model has an $8$-dimensional one.
 
We present our results in a symplectic framework. 
(The reader  is advised to consult any of the standard textbooks as \cite{AbrahamMarsden,Arnold,Trautman,Guillemin,Sternberg78}, for example.) The particular version we follow
throughout this paper, outlined in Appendix \ref{AppendixA}, is taken from Souriau's book \cite{SSD},
chapter III, pp. 123--227. The key point is that the classical motions correspond to curves (or surfaces) in an evolution space, $V$, where the dynamics takes place and is determined by a two-form $\sigma$; this can be thought of as a common generalization of both the Hamiltonian and Lagrangian formalisms.  

\goodbreak 
%%%%%%%%%%%%%%%%%%%%%%%%%%%%%%%%%%%%%%%%%%%%%%%%%%%%%%%%%%%%%%%%%%%%%%%%%%%%%%
\section{Symplectic description of the chiral model}\label{sympchirSec}
%%%%%%%%%%%%%%%%%%%%%%%%%%%%%%%%%%%%%%%%%%%%%%%%%%%%%%%%%%%%%%%%%%%%%%%%%%%%%% 

Let us assume, for simplicity that we work in a Lorentz frame where the external field is stationary. 
Variation of the chiral action (\ref{chiract}) yields the equations of motion for position $\bx$ and momentum $\bp\neq0$ in three-space, 
\beq
\left\{\barraynb{lll}
\fm 
\,\displaystyle\frac{d{\bx}}{dt}&=&\hbp+e\,{\bE}\times\bTheta+(\bTheta\cdot\hbp)\,e\,\bB,
\\[12pt]
\fm 
\,\displaystyle\frac{d{\bp}}{dt}&=&e\,\bE+{e}\,{\hbp}\times\bB+e^2(\bE\cdot\bB)\,\bTheta,
\earraynb\right.
\label{chireqmot}
\eeq
 where  $\bE$ and $\bB$ are the electric and magnetic field, respectively, and
\beq
\fm= 1+e\,\bTheta\cdot\bB
\label{chireffmass}
\eeq
is an \emph{effective mass}.   

Alternatively and equivalently, 
the chiral model (\ref{chiract}) can be described 
within a symplectic framework \cite{SSD,AbrahamMarsden,Arnold,Trautman,Guillemin,Sternberg78},  outlined in Appendix A.
For chiral fermions, the {evo\-lution space}
is 
\beq
V^7=T({\bbR}^3\!\setminus\!\{0\})\times {\bbR}
\label{V7def}
\eeq
described by triples $(\bx,\bp,t)$,
and is endowed with the two-form $\sigma$ in
(\ref{Souriauform}), i.e., 
\begin{equation}
\sigma=\omega-dh\wedge{}dt,
\label{sigma}
\eeq
specialized to the case where
\beqa
\omega&=&\omega_0+\frac{e}{2}\epsilon_{ijk}{}B^i\,dx^j\wedge{}dx^k,
\label{Ssigma}
\\[6pt]
\omega_0&=&dp_i\wedge dx^i-\frac{s}{2|\bp|^3}{}
\epsilon^{ijk}\,p_i\,dp_j\wedge dp_k,
\label{freechirsymp}
\\[6pt]
h&=&{|\bp|+e\phi},
\label{phih}
\eeqa 
where $s=1/2$
%%%%%%
\footnote{The symplectic form $\omega$ in (\ref{Ssigma}) is the sum of the free form 
$\omega_0$ (\ref{freechirsymp}) with the magnetic term,
and the Hamiltonian, (\ref{phih}), is the addition of the scalar potential to the free expression. This form is limited to time-independent fields. However, choosing instead  
$$
\sigma=\omega_1 - dh_0\wedge dt,
\quad
\omega_1=\omega_0+\frac{e}2\epsilon_{ijk}B^k dx^i\wedge dx^j + e E_i dx^i \wedge dt,
\quad
h_0=|\bp|
$$
would accommodate time-dependent fields, though \cite{SSD}. When the electric field is stationary and curl-free, $\sigma$ clearly takes
the form (\ref{sigma}). 
The subtle novelty of Souriau's prescription  was emphasised by Sternberg \cite{Sternberg78}.}.
The two-forms $\omega$ and thus $\sigma$ are closed since  
$\bnabla_{\bx}\cdot\bB=0$, and $\bnabla_{\bp}\cdot\bTheta=0$, see~(\ref{Berryterm}). 
(Remember that the points such that $\bp=0$, where the divergence of $\Theta(\bp)$
would be a Dirac delta-function, do not belong to our manifold, $V^7$.)
Wherever 
\beq
\det(\omega_{\alpha\beta})=\fm^2\equiv (1+e\,\bTheta\cdot\bB)^2\neq0,
\label{detomega}
\eeq
 the  kernel of~$\sigma$ is $1$-dimensional, and
a curve 
$(\bx(\tau),\bp(\tau),t(\tau)\big)$ is tangent to it  iff
 the equations of motion~(\ref{chireqmot}) are satisfied \footnote{The equations of motion~(\ref{chireqmot}) are also Hamiltonian, with  Hamiltonian $h$ as in (\ref{freechirsymp})  and fundamental Poisson brackets given by \cite{DHHMS}
 $$ 
\{x^i,x^j\}=\frac{-\epsilon^{ijk}\,\Theta_k}{1+e\,\bTheta\cdot\bB},
\quad
\{p_i,p_j\}=\frac{e\,\epsilon_{ijk}\,B^k}{1+e\,\bTheta\cdot\bB},
\quad
\{p_i,x^j\}=\frac{\delta_i^j+e\,\Theta_iB^j}{1+e\,\bTheta\cdot\bB}.
$$ 
It follows that the coordinates  do not commute, let alone in the free case.
}.
At points where $\det(\omega_{\alpha\beta})=0$  the system is degenerate, necessitating symplectic alias Faddeev-Jackiw \cite{FaJa} reduction.  

A constant $\bTheta$ aligned in the $z$-direction would correspond to the planar case studied in \cite{Peierls,AnAn,HMS,ZH-chiral}. Then, the vanishing of the analogous determinant, interpreted as the vanishing of an effective mass, merely requires fine-tuning of the magnetic field; the dynamical degrees of freedom drop from $4$ to $2$, and the only allowed motions are those which follow the Hall law \cite{Peierls,AnAn,HMS,ZH-chiral}. 
In the chiral case here, instead, $\bTheta$ is parallel to the momentum, $\bp$. The determinant~(\ref{detomega}) can only vanish at particular singular points of phase space, since $\bTheta=\bTheta(\bp)$ and $\bB=\bB(\bx)$. The vanishing of $\fm$ is therefore rather spurious even at such exceptional points,  since it requires the magnetic field to be of the order of the squared momentum, which appears inconsistent with the assumed adiabaticity.
 
Returning to the general case $\fm\neq0$, eqns (\ref{chireqmot}) exhibit the so-called \emph{anomalous velocity} terms which have been recognized as the main reason  behind \emph{transverse shifts} or \emph{side jumps} in spin-Hall-type effects \cite{Karplus,AHE,OHE}. Let us underline the strong similarities of the chiral system with massive semiclassical models \cite{Niu,Liouville,DHHMS} as well as with their planar counterparts \cite{Peierls,AnAn,HMS,Gieres}.
Recent study indicates that chiral fermions follow
a similar pattern and exhibit, in particular, an \emph{Anomalous Hall effect} \cite{ZH-AHE}.

%%%%%%%%%%%%%%%%%%%%%%%%%%%%%%%%%%%%%%%%%%%%%%%%%%%%%%%%%%%%%%%%%%%%%%%%%%%%%%
%%%%%%%%%%%%%%%%%%%%%%%%%%%%%%%%%%%%%%%%%%%%%%%%%%%%%%%%%%%%%%%%%%%%%%%%%%%%%%
\section{Massless spinning particles}\label{masslessSec}
%%%%%%%%%%%%%%%%%%%%%%%%%%%%%%%%%%%%%%%%%%%%%%%%%%%%%%%%%%%%%%%%%%%%%%%%%%%%%%
%%%%%%%%%%%%%%%%%%%%%%%%%%%%%%%%%%%%%%%%%%%%%%%%%%%%%%%%%%%%%%%%%%%%%%%%%%%%%%

Now we consider instead a \emph{free relativistic massless spinning particle} that we describe,
following \cite{SSD}, by a $9$-dimensional evolution space $V^9$  as follows. (See Appendix~\ref{AppendixB} for a overview of the model.)
We start  with three four-vectors $R,I,\,J$ in Minkowski space-time~$\bbR^{3,1}$ with signature  $(-,-,-,+)$.
Then we put
\begin{equation}
V^9=\left\{
R,I,J\in{\bbR}^{3,1}\,\strut\Big\vert\,I_\mu{}I^\mu=J_\mu{}J^\mu=0, I_\mu J^\mu{}=-1
\right\}
\label{evolutionspace}
\end{equation}
with $I$ future-directed.
Thus $I$ and $J$ are lightlike (nonzero) vectors gene\-rating a null $2$-plane while $R$ represents a space-time event. This particular evolution space is  obtained from the Poincar\'e group by factoring out a suitable internal $\SO(2)$ subgroup (cf. Appendix \ref{AppendixB}), and carries therefore a natural action of the Poincar\'e group.

 An equivalent, but for our purposes more convenient, description of $V^9$ uses the spin tensor. Renaming $P=I$ (which will be later interpreted as the linear momentum) the latter is defined as, 
\beq
S_{\mu\nu}=-s\,\epsilon_{\mu\nu\rho\sigma}\,P^\rho{}J^\sigma.
\label{spintensor}
\eeq 
The spin tensor satisfies $\half{S}_{\mu\nu}\,{S}^{\mu\nu}=s^2$, where $s\neq0$ is the scalar  spin (whose sign is called helicity). The condition 
\beq
S_{\mu\nu}P^\nu=0
\label{PauliLubanski}
\eeq
is plainly satisfied.
Identifying the tensor~$S=(S_{\mu\nu})$ with an element of the Lorentz Lie algebra $\ort(3,1)$, the evolution space (\ref{evolutionspace}) can also be presented as
\beq
V^9=\left\{
R,P\in{\bbR}^{3,1}, S\in\ort(3,1)\,\strut\Big\vert\,P_\mu{}P^\mu=0,\;S_{\mu\nu}P^\nu=0,\;\half{S}_{\mu\nu}{S}^{\mu\nu}=s^2\right\},
\label{spinevspace}
\eeq
with, again, $P$ future-pointing. The evolution space $V^9$ depicted on Fig. \ref{figV9} is endowed with the closed two-form
borrowed from \cite{SSD}, namely \footnote{More generally \cite{SSD} the first term in (\ref{freespinsigma}) could have a coefficient $\eta=\pm1$ which represents the \emph{sign of the energy}. In this paper, we consider  positive  energy only, $\eta=1$.},
\begin{equation}
\sigma
=
-dP_\mu\wedge{}dR^\mu
-\frac{1}{2s^2}\,d{S}^\mu_{\;\lambda}\wedge{S}^\lambda_{\;\rho}\,d{S}^\rho_{\;\mu}.
\label{freespinsigma}
\end{equation}
 The dynamics is given by the  foliation whose leaves are tangent to the kernel of $\sigma$ in~$V^9$; a world-sheet [or world-line] of the system is obtained by projecting a leaf of  the latter to Minkowski space-time, yielding its corresponding space-time track.
Calculating the kernel of (\ref{freespinsigma}) using also the constraints which define the evolution space, readily shows that a curve $(R(\tau),P(\tau),S(\tau))$  in $V^9$ 
(where $\tau$ is a real parameter) is tangent to $\ker\sigma$ iff
\begin{equation}
\left\{
\begin{array}{rcl}
P_\mu{}\dot{R}^\mu&=&0,
\\
\dot{P}^\mu&=&0,
\\
\dot{S}^{\mu\nu}
&=& P^\mu\dot{R}^\nu-P^\nu\dot{R}^\mu,
\end{array}
\right.
\label{V0ker}
\end{equation}
where  the ``dot'' stands for $d/d\tau$.
The space-time ``velocity'', $\dot{R}$, associated to any such curve is hence orthogonal to the momentum $P$. Indeed, the distribution defined by equations (\ref{V0ker})  can be integrated using space-time vectors~$Z$ orthogonal to $P$,
$P_\mu{}Z^\mu=0$, namely as,
\beq 
R^\mu\to R^\mu+Z^\mu,
\qquad
P^\mu\to P^\mu,
\qquad
S^{\mu\nu}
\to
S^{\mu\nu}+(P^\mu Z^\nu-P^\nu Z^\mu). 
\label{Zshift}
\eeq

Any  point in a leaf of
$\ker\sigma$ can be reached by choosing a suitable vector $Z$.  
Therefore at each point of $V^9$ the kernel of the two-form $\sigma$ is \emph{$3$-dimensional} and projects to space-time, according to (\ref{V0ker}), as an affine subspace of ${\bbR}^{3,1}$, spanned by all vectors at $R$ orthogonal to the linear momentum $P$.
Thus the motions of a  free massless spinning particle take place on a $3$-dimensional wave-plane, tangent to the light-cone at each space-time event $R$: the particle is \emph{not localized} in space-time \cite{Penrose67,SSD}.

 We insist that all curves which lie in a leaf
should be considered to be the same motion, left invariant by  a $Z$-shift in~(\ref{Zshift}). 
 The space of motions is the collection $M^6=V^9/\ker\sigma$ of those leaves and inherits the structure of a \emph{$6$-dimensional} symplectic manifold (see below). As we  explain it below,  spin is responsible for the space-time delocalization of massless particles. 
 %%%%%%%%%
\begin{figure}[h]
\begin{center}\vskip-5mm
\includegraphics[scale=.55]{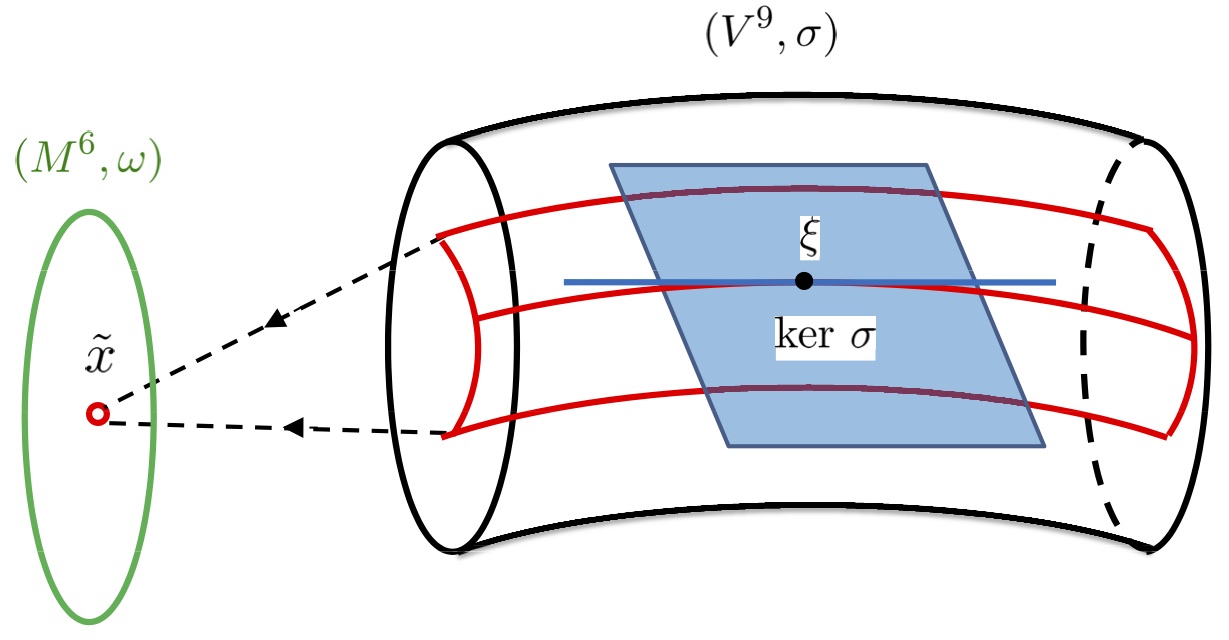}
\end{center}\vskip-8mm
\caption{\it A free massless spinning particle has a 9-dimensional evolution space $V^9$; its dynamics is defined
by a two-form $\sigma$ (see (\ref{freespinsigma})), whose kernel, $\ker\sigma$, is $3$-dimensional. A motion is a leaf tangent to the latter.
All points of a leaf can be reached by a Z-shift (also called a Wigner translation). 
The set of leaves  forms the
space of motions, $M^6$, whose points are $\tilde{x}=(\tilde{\bx},\tilde{\bp})$.
} 
\label{figV9}
\end{figure}
%%%%%%%%%
  
To obtain down-to-earth expressions, we put
$ 
R=(\br,t)
$
where $\br$ and $t$ are the position and time coordinates in a chosen Lorentz frame. The two null-vectors are in turn
$P=(\bp,|\bp|)$ and $J=(\bq,-|\bq|)$, 
where $\bp$ and $\bq$ are two (necessarily nonzero) $3$-vectors which satisfy $\bp\cdot\bq+|\bp|\,|\bq|=1$ by (\ref{evolutionspace}). In these terms we have
\beqa
S_{ij}=\epsilon_{ijk}\,s^k,
\qquad
\bs= s\big(\bp|\bq|+\bq|\bp|\big),
\qquad
S_{j4}=s\big(\bp\times\bq\big)_j=
\big(\hbp\times\bs\big)_j.
\label{Spq}
\eeqa 

We label each leaf  of $\ker\sigma$ by picking a representative point in each of them in a way which is convenient for our purposes. To this end, we first observe that
$
\tau\to(R+\tau P,P,S)
$
is an integral curve of  $\ker\sigma$ for any
given $(R,P,S)$, i.e., a particular ``motion''.
Next, shifting this curve 
by 
\beq 
Z=
\big(\frac{\hbp}{|\bp|}\times\bs,0\big)
\label{goodshift}
\eeq 
yields another integral curve lying in the same leaf. 
Finally, taking $\tau=-t/|\bp|$
yields the point which has
zero time coordinate; this is the point that we choose. 
See Fig. \ref{figZshift} to illustrate our strategy.
%%%%%%%%%%%%%
\begin{figure}[ht]
\begin{center}
\includegraphics[scale=.6]{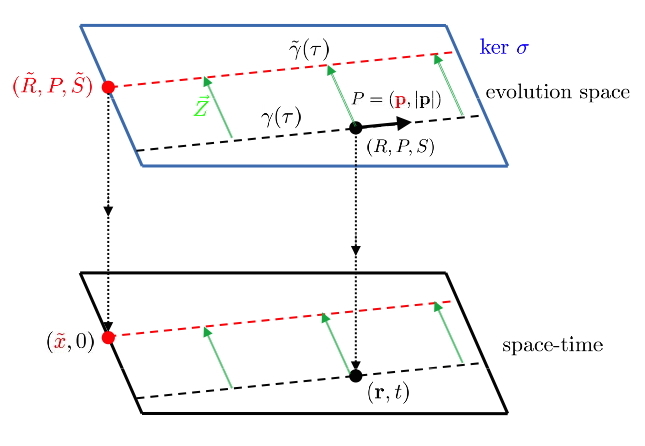}
\vspace{-12mm}
\end{center}
\caption{\it The spin of a motion tangent to  the ($3$-dimensional) kernel of $\sigma$ can be enslaved by a suitable Z-shift.  Choosing the point 
$\red{(\tilde{R},P,\tilde{S})}$
on the shifted curve with vanishing  time coordinate, \red{$\tilde{R}=(\tilde{\bx},0)$},
 provides us with a coordinate $\red{\tilde{\bx}}$ of  the motion-sheet. The characteristic leaves in $V^9$ project to Minkowski space-time as $3$-planes orthogonal to the momentum: a massless spinning particle cannot be localized.
} 
\label{figZshift}
\end{figure}
%%%%%%%%%
The corresponding point on the shifted curve has position 
$R=(\tbx,0)$.  
Spin becomes enslaved to the linear $3$-momentum,
\beq
S_{j4}=0, \qquad
\bs=s\,{\hbp}.
\label{enslavement}
\eeq
An important observation which follows from (\ref{Spq}) is that 
\beq
\hbp\cdot\bs=s
\label{sproj}
\eeq
 in general, and not only in 
the enslaved case (\ref{enslavement}). It is thus \emph{not} the length of the $3$ vector~$\bs$ but its
\emph{projection} along $\hbp$  which is a constant.

In terms of $(3+1)$-variables, $Z=(\bZ,\hbp\cdot\bZ)$
 the ``Z-shift'' (\ref{Zshift}) acts as
\beq
\br\to\br+\bZ,
\qquad
t\to t+\hbp\cdot\bZ,
\qquad
\bp\to\bp,
\qquad
\bs\to\bs+\bp\times\bZ.
\label{Zshiftps} 
\eeq
Thus, in the free case, the freedom of Z-shifting allows us to \emph{eliminate the spin as an independent degree of freedom} altogether and the entire leaf can be labelled by
$ \tbx$ and $\tbp=\bp\neq0$ alone.
The latter parametrize the space of motions $M^6=V^9/\ker\sigma$, which has therefore  the topology of $T({\bbR}^3\!\setminus\!\{0\})$. 
Finally, the two-form $\sigma$ in~(\ref{freespinsigma}) descends to 
the space of motions $M^6$ as the symplectic two-form 
 (\ref{freechirsymp}), namely
\beq
\omega
=d\tilde{p}_i\wedge d\tilde{x}^i-\frac{s}{2|\tbp|^3}\,\epsilon^{ijk}\, 
\tilde{p}_i\,d\tilde{p}_j\wedge d\tilde{p}_k.
\label{freem0shalf}
\eeq

%%%%%%%%%%%%%%%%%%%%%%%%%%%%%%%%%%%%%%%%%%%%%%%%%%%%%%%%%%%%%%%%%%%%%%%%%%%%%%
%\subsection{Poincar\'e symmetry}\label{Psymm}
%%%%%%%%%%%%%%%%%%%%%%%%%%%%%%%%%%%%%%%%%%%%%%%%%%%%%%%%%%%%%%%%%%%%%%%%%%%%%%

Now  we establish the Poincar\'e symmetry of the model.
 The Poincar\'e Lie algebra $\se(3,1)$, spanned by the pairs $(\Lambda, \Gamma)$ where $\Lambda=(\Lambda_{\mu\nu})$ belongs to the Lorentz Lie algebra $\sort(3,1)$, and $\Gamma=(\Gamma^\mu)$ is a translation in Minkowski space-time, ${\bbR}^{3,1}$, acts on $V^9$ by the  lift of its action on Minkowski-space-time. 
This action on $V^9$ reads as follows
\beq
\delta R^\mu = \Lambda^{\mu}_{\,\nu}R^\nu+\Gamma^\mu,
\qquad
\delta P^{\mu}=\Lambda^{\mu}_{\;\nu}P^\nu,
\qquad
\delta S^{\mu\nu}=\Lambda^\mu_{\,\rho}S^{\rho\nu}
-
\Lambda^\nu_{\,\rho}S^{\rho\mu},
\label{spinactonV}
\eeq
and clearly leaves the two-form (\ref{freespinsigma}) invariant. It is therefore a symmetry of the system, which descends  to the space of motions $(M^6,\omega)$.  The associated Noetherian conserved quantities are 
\begin{equation}
P^\mu=I^\mu,
\qquad
M^{\mu\nu}=R^{\mu} P^\nu-R^\nu P^{\mu}+S^{\mu\nu},
\label{muPoincare}
\end{equation}
which identifies the vector $P$ and the bi-vector $M$ as the \emph{conserved linear and angular momentum}, respectively.

To get explicit formulas in a $(3+1)$-decomposition, we parametrize the Poincar\'e Lie algebra by
$\Lambda_{ij}=\epsilon_{ijk}\,\omega^k$,
$\Lambda_{i4}=\beta^i$ 
and $\Gamma=(\bgamma,\varepsilon)$,
where $\bomega,\bbeta,\bgamma\in\bbR^3,\varepsilon\in\bbR$ are infini\-tesimal rotations, boosts and space- and time-translations, respectively. In terms of this decomposition, the infinitesimal Poincar\'e-action on $V^9$ is given, see (\ref{spinactonV}) and (\ref{Spq}), by
\beq
\left\{\barraynb{lll}
 \delta\br&=&\bomega\times\br+\bbeta{}t+\bgamma,
\\
\delta t&=&\bbeta\cdot\br+\varepsilon,
\\
\delta\bp&=&\bomega\times\bp+\bbeta\,|\bp|,
\\
\delta\bs&=& \bomega\times\bs-\bbeta\times(\hbp\times\bs),
\earraynb\right.
\label{infPVrtps}
\eeq 
and duly projects to Minkowski space-time as the natural one.

To write down the explicit form of the Poincar\'e momenta (\ref{muPoincare}) we present the 
 matrix $M=(M_{\mu\nu})$ (which belongs to the dual of the Lorentz algebra) as $M_{ij}=\epsilon_{ijk}\,\ell^k$ 
and $M_{j4}=g^j$ with $\belle$ and $\bg$ two $3$-vectors.
 In terms of the above $(3+1)$-parametrization we find 
\beq
\left\{\barraynb{lll} 
\belle&=&
 \br\times\bp\,+\bs,
\\[4pt] 
\bg&=&
 \np\,\br-\bp t+\hbp\times\bs.
\earraynb\right.
\eeq
Then the quantity
\beq 
\tbx=\frac{\bg}{|\bp|}=
 \br-\hbp\,{}t+\frac{\hbp}{|\bp|}\times\bs
\label{txr}
\eeq
is itself conserved. Working out the action of the full Poincar\'e Lie algebra (\ref{spinactonV}) 
on the space of motions~$(M^6,\omega)$ provides us with \footnote{One of us (CD) discussed the infinitesimal action (\ref{DuvalZiegler}) and the quantities listed in (\ref{Table}) with F. Ziegler long ago (unpublished).} 
\begin{eqnarray}\left\{\barraynb{lll}
\delta\tbp&=&\bomega\times\tbp+\,|\tbp|\,\bbeta,
\\[6pt]
\delta\tbx&=&\bomega\times\tbx+
s\,\bbeta\times\displaystyle\frac{\tbp}{|\tbp|^2}
-\bbeta\cdot\tbx\,\displaystyle\frac{\tbp}{|\tbp|}+\bgamma-\varepsilon\,\frac{\tbp}{|\tbp|}.
\earraynb\right.
\label{DuvalZiegler}
\end{eqnarray}

The  $10$-parameter vector field (\ref{DuvalZiegler}) leaves the free symplectic structure (\ref{freem0shalf}) invariant, i.e., it generates a family of symmetries, to which the symplectic Noether theorem \cite{SSD} associates  $10$ constants of the motion, namely
\beq
\left\{
\barraynb{lllll}
\belle&=&\tbx\times\tbp+s{\hbp}&\qquad
&\hbox{angular momentum}
\\
\bg&=&|\tbp|\,\tbx &&\hbox{boost momentum}
\\
\bp&=&\tbp  &&\hbox{linear momentum}
\\
\cE&=&|\tbp| &&\hbox{energy}
\earraynb\right.
\label{Table}
\eeq
whose conservation follows also directly from the free equations of motions. Note that the two terms in the free angular momentum  $\belle$
are separately conserved 
%%%%%%%%%%%%%%%
\footnote{
 This is unlike to the Dirac equation, where internal and external degrees of freedom are intricately mixed, due to the need of using Dirac matrices to represent the Lorentz group. Our classical model might comply with squaring the Dirac equation, allowing indeed to separate the orbital and spin terms.}.
%%%%%%%%%%%%%%%%
  
The Poisson brackets of the quantities in (\ref{Table}) calculated using (\ref{freem0shalf}), 
\beq
\barraynb{lllll}
\{\ell_i,\ell_j\}=-\epsilon_{ij}^k\,\ell_k,\qquad
&\{\ell_i,g_j\}=-\epsilon_{ij}^k\,g_k,\qquad
&\{\ell_i,p_j\}=-\epsilon_{ij}^k\,p_k,\qquad
&\{\ell_i,\cE\}=0,\qquad&
\\[6pt]
\{g_i,g_j\}=\epsilon_{ij}^k\,\ell_k,
&\{g_i,p_j\}=-\cE\,\delta_{ij},
&\{g_i,\cE\}=-p_i,
&\{p_i,p_j\}=0,\quad
&\{p_i,\cE\}=0,
\earraynb
\label{Palg}
\eeq
are those of the \emph{Poincar\'e Lie algebra} $\se(3,1)$, as expected. Calculating the Casimir invariants 
\beq
m^2=-\bp^2+\cE^2=0,
\qquad
\belle\cdot\hbp= s,
\label{casimirs}
\eeq
shows that the infinitesimal Poincar\'e symmetry we have just found  is realized in the \emph{zero-mass and spin-$s$ representation}. 

The reason hidden behind all this is that the (connected) Poincar\'e group acts on the space of motions  symplectically and transitively. Therefore~$(M^6,\omega)$ is  a \emph{coadjoint orbit of the Poincar\'e group} \cite{SSD}.  
The symplectic form (\ref{freem0shalf}) is, in particular, Souriau's $\# (17.145)$ in \cite{SSD}. 
The $Z$-translations in equation (\ref{Zshift}), also  identified as Wigner translations \cite{Wigner}, belong to the stability subgroup, $\SO(2)\times{\bbR}^3$, of the Poincar\'e-action of a basepoint in the orbit.  

So far, we have considered the infinitesimal action of the Poincar\'e \emph{Lie algebra}. The construction 
allows us to  work out the \emph{finite action} of 
the \emph{connected} (also called neutral) 
Poincar\'e \emph{Lie group} on the space of motions $(M^6,\omega)$. To that end it is enough to spell out its natural action 
\beq
(R,I,J)\to(R'=\cL{}R+\cC,I'=\cL{}I,J'=\cL{}J)
\eeq
 with $\cL\in\SO_+(3,1)$ and $\cC\in{\bbR}^{3,1}$, integrating the infinitesimal action (\ref{spinactonV}) on the evolution space~$V^9$ introduced in (\ref{evolutionspace}).
Parametrizing the connected Poincar\'e group by $A$ (rotation), $\bb=b\,\bu$ (boost in the direction~$\bu$),
$\bc$ (space-translation), and $e$ (time-translation), a
 tedious calculation summarized in  Appendix \ref{AppendixC}) yields the  action $(\tbp,\tbx)\to(\tbp',\tbx')$, where
\beq
\left\{
\barraynb{lll}
\tbp' &=& {A}\,\tbp + (\gamma-1)\, (\bu\cdot{A}\,\tbp)\,\bu + \gamma\, |\tbp|\, \bb,
%\label{Ponp}
\\[12pt]
\tbx' &=& \displaystyle\frac{1}{|\tbp|+\bb\cdot{A}\,\tbp}\,\bigg[
\bb\times{A}\big(\tbx\times\tbp+s\,\displaystyle\frac{\tbp}{|\tbp|}\big)
\\[12pt]
&&
+\,|\tbp|\,A\,\tbx+(\gamma-1)\,|\tbp|\,(\bu\cdot{A}\,\tbx)\,\bu\,
-\,\gamma\,|\tbp|(\bb\cdot{A}\,\tbx)\,\bb+(\bb\cdot {A}\,\tbp)\,\bc
\\[6pt]&&
-\displaystyle\frac{e}{\gamma}\Big({A}\,\tbp+(\gamma-1)\, (\bu\cdot{A}\,\tbp)\,\bu\Big)
+\,|\tbp|\,(\bc-\bb\,{e}) 
\bigg],
%\label{Ponx}
\earraynb
\right.
\label{Ponpx}
\eeq
with $\gamma=(1-\vert\bb\vert^2)^{-1/2}$ as usual; by keeping ``tildes" we insist that our variables live on the space of motions (remember that  $\tbp=\bp$ but $\tbx\neq\br$). This  extends the
infinitesimal action  (\ref{DuvalZiegler}) to finite transformations.

In a Lorentz frame the trajectory labeled with  $\tbx$ is given by (\ref{txr}), i.e.,
\beq
\tbx=\br-\hbp\,{t}+\frac{\hbp}{|\bp|}\times\bs,
\label{trajectory}
\eeq
which describes motion  with the velocity of light, directed along $\hbp$.  A Z-shift displaces the trajectory;
starting, in particular, with ``enslaved'' spin, the latter is  ``unchained'' and the trajectories one obtains fill a three-plane in $4$-space. However, it is easy to see using (\ref{infPVrtps}) that the right hand side
 of (\ref{trajectory}) remains invariant: the motion is not affected.
 
Intuitively, the freedom of Z-shifting is reminiscent of \emph{gauge freedom}: it can always be performed at will; enslaving spin is in turn a sort of \emph{gauge fixing}, allowing to interpret the result in terms of physical degrees of freedom alone.

We  mention for completeness that the Poincar\'e symmetry of the massless spinning particle actually extends 
to an $\sort(4,2)$ conformal symmetry. See, e.g., \cite{CFH}.

%%%%%%%%%%%%%%%%%%%%%%%%%%%%%%%%%%%%%%%%%%%%%%%%%%%%%%%%%%%%%%%%%%%%%%%%%%%%%%
%%%%%%%%%%%%%%%%%%%%%%%%%%%%%%%%%%%%%%%%%%%%%%%%%%%%%%%%%%%%%%%%%%%%%%%%%%%%%%
\section{Poincar\'e symmetry of the free chiral model}\label{chirsymm}
%%%%%%%%%%%%%%%%%%%%%%%%%%%%%%%%%%%%%%%%%%%%%%%%%%%%%%%%%%%%%%%%%%%%%%%%%%%%%%
%%%%%%%%%%%%%%%%%%%%%%%%%%%%%%%%%%%%%%%%%%%%%%%%%%%%%%%%%%%%%%%%%%%%%%%%%%%%%%

Now we return to chiral fermions. Does the free system (\ref{chiract}) admit a Poincar\'e symmetry~? 
For $\bE=\bB=0$ the motions can be determined explicitly: the $\bTheta$-term drops out from  (\ref{chireqmot}), yielding
\beq  
\bx(t)=\tbx+\hbp{\,}t,
\qquad 
\bp(t)=\tbp,
\eeq 
with $\tbx$ and $\tbp$ constant vectors (and $\hbp=\tbp/|\tbp|$). 
As explained in Sect. \ref{sympchirSec},
the  chiral space of motions  
 $M^6=V^7/\ker\sigma$ can, therefore, be 
 labeled
by the constants of the motion
\beq
\tbx=\bx(t)-\hbp\,{t}
\quad\hbox{and}\quad
\tbp.
\eeq
With the fields switched off, 
  the \emph{two-form~$\omega$ in (\ref{Ssigma}) becomes precisely (\ref{freem0shalf})}:  \emph{the free chiral model has the same space of motions as that of the massless spinning particle with $s=1/2$}, studied in Section \ref{masslessSec}.
  
Then, our strategy is to ``import''  the natural Poincar\'e symmetry of the massless spinning model to the chiral system  through their common space of motions.
From the identity of  the space-of-motions coordinates $(\tbx,\tbp)$ we conclude that,
in terms of the coordinates $(\bx,\bp,t)$ on the chiral evolution space 
$V^7$, the strange-looking  Poincar\'e  action (\ref{DuvalZiegler})  (with $s=1/2$) becomes,
\begin{eqnarray}
\label{PConV7}
\left\{\barraynb{lll}
\delta\bx&=&
\bomega\times\bx+\displaystyle\bbeta\times\frac{\hbp}{2|\bp|}+\bbeta\,t+\bgamma,
\\
\delta\bp&=&\bomega\times\bp+|\bp|\,\bbeta,
\\
\delta{t}&=&\bbeta\cdot\bx+\varepsilon.\quad
\earraynb\right.
\end{eqnarray}
By construction, these vector fields generate the same Lie algebra as those in (\ref{DuvalZiegler}), namely the Poincar\'e algebra $\se(3,1)$.

Equation (\ref{PConV7}) confirms  the recently proposed action of the Lorentz subalgebra on chiral fermions \cite{ChenSon}. 
The conserved quantities associated with the  generators  of the latter are, in particular,
\begin{eqnarray}
\left\{\barraynb{lll}
\belle&=&\bx\times\bp\,+\half\,\hbp,
\\
\bg&=&|\bp|\,\bx-\bp{}t,
\earraynb\right.
\end{eqnarray}
as it can be checked directly by showing that the infinitesimal rotations and boost generators in equation (\ref{PConV7}) Lie-transport the two-form 
$\sigma$ in (\ref{sigma})--(\ref{freechirsymp}), and then by calculating the associated Noetherian quantities.

We have thus established a twisted Poincar\'e symmetry of the free chiral system. We insist, however, that the action  (\ref{PConV7}) is \emph{not} the usual, natural one on Minkowski space-time.  In fact, it is \emph{not} an
action on space-time at all, since it also involves the momentum variable~$\bp$; 
it is rather a sort of dynamical symmetry --  but one for the free dynamics. 

In conclusion, the  chiral model \emph{admits a  Poincar\'e  symmetry, but, unlike for the massless spinning model, this  symmetry does not act in the usual, natural way}.  It follows   that $\bx$ should \emph{not} be considered 
as a \textit{bona fide} position variable, because it \emph{does not transform} under a boost as positions should~: it labels  a \emph{motion} and is not a space coordinate.  We contend that the \textit{well-founded} position of our particle should rather be regarded as given by the three spatial coordinates, $\br$, relatively to a chosen Lorentz frame, of intrinsic space-time translations within the Poincar\'e group. 
 From the identity of  the space-of-motions coordinates $(\tbx,\tbp)$ we conclude in fact that the coordinates, $\bx$, of  the chiral particle are related 
 to those, $\br$, of the massless Poincar\'e model according to
\beq
 \bx=\br+\frac{\hbp}{|\bp|}\times \bs 
 \qquad\hbox{with}\qquad
 \bs\cdot\hbp=\half.
\eeq
The coordinates coincide,
$\bx=\br$, only when spin is enslaved.

%%%%%%%%%%%%%%%%%%%%%%%%%%%%%%%%%%%%%%%%%%%%%%%%%%%%%%%%%%%%%%%%%%%%%%%%%%%%%%
%%%%%%%%%%%%%%%%%%%%%%%%%%%%%%%%%%%%%%%%%%%%%%%%%%%%%%%%%%%%%%%%%%%%%%%%%%%%%%
\section{Coupling to an external electromagnetic field}\label{emcouplingSec}
%%%%%%%%%%%%%%%%%%%%%%%%%%%%%%%%%%%%%%%%%%%%%%%%%%%%%%%%%%%%%%%%%%%%%%%%%%%%%%
%%%%%%%%%%%%%%%%%%%%%%%%%%%%%%%%%%%%%%%%%%%%%%%%%%%%%%%%%%%%%%%%%%%%%%%%%%%%%%

Let us now cope with a number of procedures enabling us to couple our relativistic massless and spinning particle to an external electromagnetic field.

The conventional minimal coupling rule says that
the $4$-momentum should be shifted by the $4$-potential as follows,
\beq
p_\mu\to p_\mu-eA_\mu.
\label{usualcoupling}
\eeq
This is \emph{not exactly} what is proposed in (\ref{chiract}), though: while the rule
(\ref{usualcoupling}) is used for the $4$-momentum $(\bp,h)$, the $3$-vector $\bp$ in the Berry term $\bTheta$ is \emph{not} shifted. 
Remarkably, 
this ``half-way-rule'' is instead consistent with working with the same evolution space as for a free particle, but adding the electromagnetic field strength $F$ to the free two-form (\ref{freespinsigma}) \cite{SSD}, 
\beq
\sigma\to\sigma+eF,
\label{Scoupling}
\eeq 
where $e$ is the electric charge of the system. This two-form is still closed, $d\sigma=0$, because~$F$ is a closed two-form of Minkowski space-time.

\goodbreak

The rules (\ref{usualcoupling}) and (\ref{Scoupling}) are equivalent  only in the spinless case. Then why should~(\ref{Scoupling}) be chosen? An argument in its favor comes form our personal experience of working in the plane, where it yielded an insight into Hall-type phenomena \cite{Peierls,AnAn,HMS,ZH-chiral,AHE,OHE}. In non-commutative mechanics in the plane, modifications of the principle 
(\ref{Scoupling}) lead  to unsatisfactory models, see~\cite{HMS,LSZ}. 
The merits of (\ref{Scoupling}) 
have  been praised, for example, by Sternberg 
\cite{Sternberg78}. It is hence this scheme we will be using throughout this paper.

%%%%%%%%%%%%%%%%%%%%%%%%%%%%%%%%%%%%%%%%%%%%%%%%%%%%%%%%%%%%%%%%%%%%%%%%%%%%%%
\subsection{Minimal coupling of the massless spinning model}\label{minimalSec}
%%%%%%%%%%%%%%%%%%%%%%%%%%%%%%%%%%%%%%%%%%%%%%%%%%%%%%%%%%%%%%%%%%%%%%%%%%%%%%

Applying the prescription (\ref{Scoupling}) to the massless spinning model of Section \ref{masslessSec} yields, on the evolution space $V^9$ in (\ref{spinevspace}), the closed two-form
\beq
\sigma=
-dP_\mu\wedge{}dR^\mu
-\frac{1}{2s^2}\,d{S}^\mu_{\;\lambda}\wedge{S}^\lambda_{\;\rho}\,d{S}^\rho_{\;\mu}
+\frac{e}2 F_{\mu\nu}\,
dR^\mu\wedge dR^\nu.
\label{mincoupsigma}
\eeq
Then a lengthy calculation using the constraints  in the definition  (\ref{spinevspace}) of $V^9$ shows 
that the equations of free motions (\ref{V0ker}) change to \footnote{
The system would become singular at those points of $V^9$ where~$S_{\alpha\beta}F^{\alpha\beta}=0$; this would change locally the dimension of $\ker\sigma$, destroying a priori the smooth manifold structure of the space of motions. 
}
\begin{equation}
\left\{
\begin{array}{rcl}
\dot{R}^\mu&=&P^\mu{}+\displaystyle\frac{S^{\mu\nu}
{F}_{\nu\rho}P^\rho}{\half{S\cdot{}F}
},
\\[8pt]
\dot{P}^\mu&=&-eF^\mu_{\;\nu}\dot{R}^\nu,
\\[8pt]
\dot{S}^{\mu\nu}
&=&
P^\mu\dot{R}^\nu-P^\nu\dot{R}^\mu. 
\end{array}
\right.
\label{VFker}
\end{equation}
assuming that ${S\cdot{}F}\equiv{S}_{\alpha\beta}F^{\alpha\beta}\neq0$.
The dimension of $\ker\sigma$  drops from $3$ to $1$: 
 the spin-field coupling
 term  in the velocity relation  breaks 
 the Z-shift-invariance. It follows that the spin degree can not now be eliminated and we are left with a \emph{$8$-dimensional space of motions} (phase space, locally).

Let us now express the equations of motion (\ref{VFker}) in terms of the $(3+1)$-decomposition we introduced in the previous section.
Assuming, that  
\beqa
(a)\quad\half{S\cdot{}F}\equiv\half{S}_{\alpha\beta}F^{\alpha\beta}=\bs\cdot(\bB-\hbp\times\bE)\neq0,
\qquad
(b)\quad
\hbp\cdot\bB\neq0,
\label{regconds}
\eeqa
a strange cancellation takes place  in 
the velocity relation in  (\ref{VFker}), which becomes
\beq
\dot{\br}=s|\bp|\displaystyle\frac{\bB-\hbp\times\bE}{\bs\cdot(\bB-\hbp\times\bE)},
\qquad
\dot{t}=
s|\bp|\displaystyle\frac{(\hbp\cdot\bB)}{\bs\cdot(\bB-\hbp\times\bE)}.
\label{rdottdot}
\eeq 
Condition (a), the  analog of the nonvanishing of the effective mass, (\ref{detomega}),  will henceforth be assumed to hold.  

Condition (b) in  (\ref{regconds}) requires that the momentum should not be perpendicular to the magnetic field. 
When it is not satisfied, then
$\dot{t}=0$, so that, while the motion still takes place along a curve, it becomes \emph{instantaneous} \footnote{Instantaneous motions with infinite  velocity are familiar in non-relativistic optics \cite{OHE}.  Intriguingly, motion with superluminal velocity also appears in certain higher-order massless relativistic models \cite{Plyush}.}. 

Let us assume that the regularity conditions (\ref{regconds}) hold; then merging the two equations in (\ref{rdottdot})  provides us with 
\begin{equation} 
\left\{
\begin{array}{rcl}
\displaystyle\frac{d\br}{dt}&=&\displaystyle\frac{\bB-\hbp\times\bE}{\hbp\cdot\bB},
 \\[12pt]
\displaystyle\frac{d\bp}{dt}&=&
 e\big(\bE+\displaystyle\frac{d\br}{dt}\times\bB\big)
 =e\,\displaystyle\frac{\bE\cdot\bB}{\hbp\cdot\bB}\,\hbp,
\\[12pt]
\displaystyle\frac{d{\bs}}{dt}
&=&\displaystyle\bp\times\frac{d{\br}}{dt}
=\displaystyle\frac{\bp\times\bB}{\hbp\cdot\bB}-\frac{\bp\times(\hbp\times\bE)}{\hbp\cdot\bB}.
\end{array}
\right.
\label{tg0eqmot}
\end{equation} 

We insist on the rather unusual form of these equations.
Firstly, the $\hbp$ one would have expected on the r.h.s. of the velocity relation cancels  out; the electric charge drops out also. The dynamics of the momentum decouples from the spin  as long as the latter does not vanish; also the scalar spin $s\neq0$ disappears  from all equations.  
Equations (\ref{tg0eqmot}) imply that $d\hbp/dt=0$ so that the direction of $\bp$ is unchanged during the motion. Spin is in fact not an independent variable, its (for space-time dynamics irrelevant) motion is entirely determined by the other dynamical data \footnote{
Somewhat surprisingly, switching off  the external fields in 
(\ref{tg0eqmot}) does not yield the free system.
Remember, though, that the equations (\ref{tg0eqmot}) are derived under the assumption that the regularity conditions (\ref{regconds}) be satisfied --- which is clearly not the case when the electromagnetic field vanishes. Accordingly, the transition from the interacting to the free case is not smooth, as highlighted by the dimension of the space of motions dropping from~$8$ to $6$.  The correct way of tackling the problem would be, once again, Faddeev-Jackiw reduction \cite{FaJa} we do not consider here.}.

Let us put, for example, our massless but charged particle into  crossed constant electric and magnetic fields  like in the Hall effect, $\bE=E\,\hat{\bx}$, $\bB=B\,\hat{\bz}$, (say). Then~$\bp$ is itself a constant of the motion, and so is the angle $\vartheta$ between $\bB$ and $\bp$ (which cannot be $\pi/2$ for~$\bp\cdot\bB\neq0$).
Let us assume for simplicity that the initial momentum lies in the $x$-$z$ plane.  
%%%%%%%%%
\begin{figure}[h]
\begin{center}\vspace{-8mm}
\includegraphics[scale=.5]{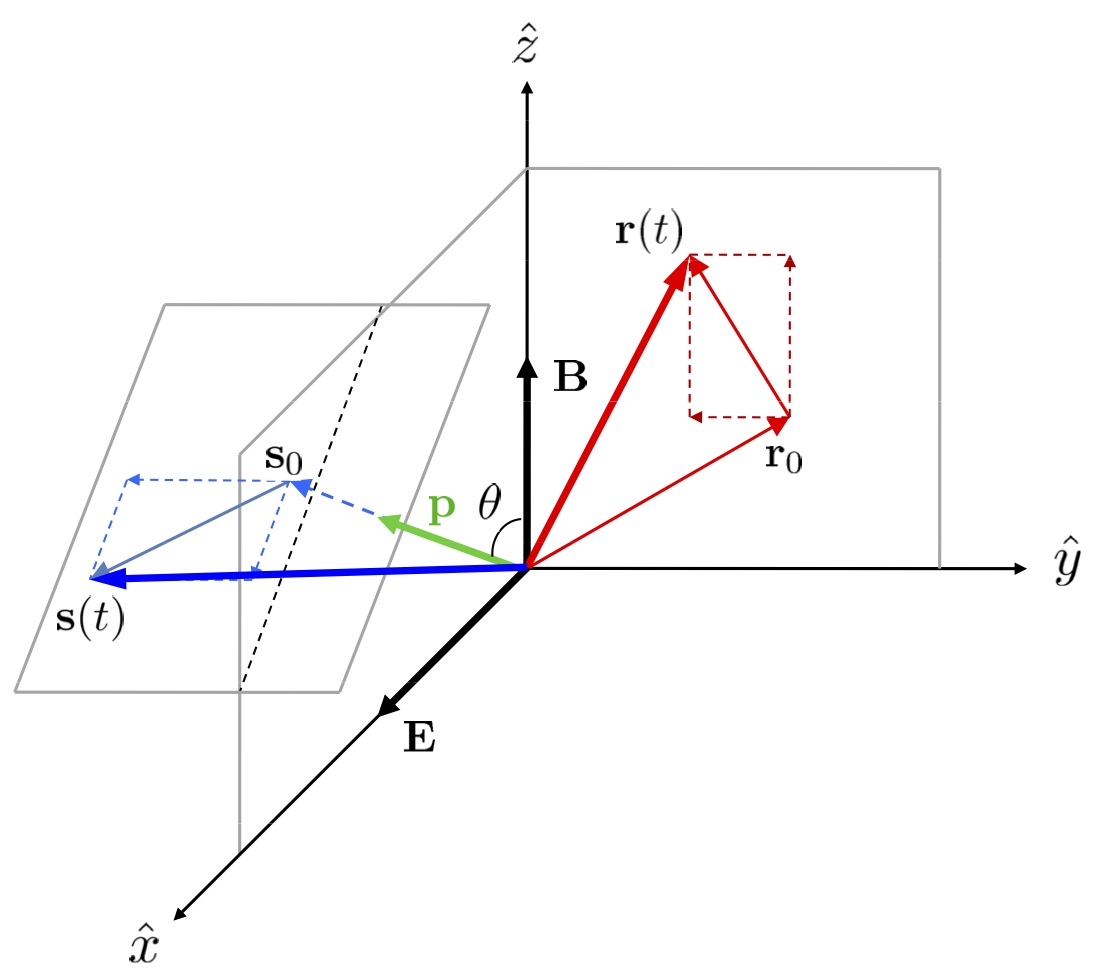}
\vspace{-9mm}
\end{center}
\caption{\it Motion in Hall-type electric and magnetic fields described by equation (\ref{g0motion}).  
The initial position, $\br_0$, is chosen in the $y$-$z$ plane, and the initial momentum and spin are chosen to be parallel and in the $x$-$z$ plane.
Then, the spatial motion, $\red{\br}(t)$,  is a combination of constant-velocity Hall drift  perpendicularly to $\bE$ and $\bB$, combined with constant-velocity vertical drift. The momentum, $\green{\bp}$, is conserved, but the spin, $\blue{\bs}(t)$, moves in a plane perpendicular to the momentum.} 
\label{g0plot}
\end{figure}
%%%%%%%%%

Then the equations of motion are solved by
\beq
\left\{\barraynb{lll}
\br(t)&=&\Big((\cos\vartheta)^{-1}\,\hat{\bz}-\displaystyle\frac{E}{B}\,\hat{\by}\Big)t
+\br_0,
\\[4pt]
\bp(t)&=&\bp_0,
\\[4pt]
\bs(t)&=& |\bp|\Big(-\tan\vartheta\,\hat{\by}+\displaystyle\frac{E}{B}\big(\cos\vartheta\,\hat{\bx}-\sin\vartheta\,\hat{\bz})
\Big)\,t+\bs_0.
\earraynb\right.
\label{g0motion}
\eeq
Thus, in addition to a constant-speed vertical motion, the ``particle'' also  drifts 
perpendicularly to the electric field  with \emph{Hall velocity}
${E}/{B}.$
The spin vector follows an even more curious motion perpendicularly to $\hbp$ so that its projection on $\hbp$ remains a constant,
$
\bs(t)\cdot\hbp=\bs_0\cdot\hbp.
$ 
Thus while spin is decoupled, it can \emph{not} consistently be enslaved as in (\ref{enslavement}) since $\bs$ and~$\hbp$ do not remain parallel
even if we start with a such initial condition, see Fig. \ref{g0plot}. Note that $S.F=2sB\cos\vartheta$ for such a motion and the system is regular therefore when $\vartheta\neq\pi/2$.

\goodbreak

 The velocity is \emph{superluminal} ($|d\br/dt|>1$) and diverges as $\vartheta\to\pi/2$; for $\hbp\cdot\bB=0$ we get instantaneous motions, i.e. with infinite velocity, parallel to the~$z$ axis. 
 This is in fact a general property, as seen from (\ref{VFker}), because
$\dot{R}_\mu \dot{R}^\mu<0$,
the $4$-vector $(S^{\mu\nu}
{F}_{\nu\rho}P^\rho)$ being space-like. 

%%%%%%%%%%%%%%%%%%%%%%%%%%%%%%%%%%%%%%%%%%%%%%%%%%%%%%%%%%%%%%%%%%%%%%%%%%%%%%%%%%%%%%%%%%%%%%%%%
\subsection{Anomalous coupling}\label{anomSec}
%%%%%%%%%%%%%%%%%%%%%%%%%%%%%%%%%%%%%%%%%%%%%%%%%%%%%%%%%%%%%%%%%%%%%%%%%%%%%%%%%%%%%%%%%%%%%%%%%

The model of Section \ref{minimalSec} is not completely satisfactory, and now we generalize our  minimal scheme. Our clue is to allow the ``mass-square'' $P_\mu{P}^\mu$ to depend on the  coupling of spin to the electromagnetic field as suggested in \cite{Souriau74,Duval75}, i.e.,
\beq
P_\mu{P}^\mu=-\frac{eg}{2}\, S\cdot{}F,
\label{gmass}
\eeq
where we used once again the shorthand
$
S\cdot{}F\equiv{}S_{\alpha\beta}F^{\alpha\beta}
$, cf. (\ref{regconds}). The
real constant $g$ will be interpreted as the \emph{gyromagnetic ratio}
\footnote{Equation (\ref{gmass}) can be generalized by putting $P_\mu{P}^\mu=f(eS.F)$ where $f$ is an otherwise arbitrary function such that $f(0)=0$. 
We refer to \cite{Souriau74,Duval75} for the case of massive spinning particles non-minimally coupled to an external electromagnetic field.}.
Generalizing the previous relation $P=I$ as
\beq
P^\mu=I^\mu+\frac{eg}{4}\,
(S\cdot{}F) 
J^\mu,
\label{animp}
\eeq
where $I$ and $J$ are still as in (\ref{evolutionspace}), helps us to implement the equation of state (\ref{gmass}). The condition
$ 
S_{\mu\nu}P^\nu=0
$ is also automatically satisfied.

\goodbreak

Hence we introduce the novel evolution space   
\begin{equation}
\tV^9=\left\{
P,R\in{\bbR}^{3,1},S\in\ort(3,1)\,\strut\Big\vert\,P_\mu{}P^\mu=
-\frac{eg}{2}\, S\cdot{}F, 
\; S_{\mu\nu}P^\nu=0,\; \half{S}_{\mu\nu}{S}^{\mu\nu}=s^2\right\},\quad
\label{gspinevspace}
\end{equation}
endowed with  the  closed two-form,
\begin{equation}
\sigma
=
-dP_\mu\wedge{}dR^\mu
-\frac{1}{2s^2}\,d{S}^\mu_\lambda\wedge{S}^\lambda_\rho\,d{S}^\rho_\mu
+\half{}eF_{\mu\nu}\,dR^\mu\wedge{}dR^\nu.
\label{gsigma}
\end{equation}
Note that (\ref{gsigma}) is formally the same as  (\ref{mincoupsigma}) up to the mass-shell constraint (\ref{gmass}).

Some more effort is needed to work our the new equations of motion from the kernel of~$\sigma$ using the  constraints which define $\tV^9$. 
We find that a curve $(R(\tau),P(\tau),S(\tau))$ is tangent to $\ker\sigma$ in (\ref{gsigma}) iff
\begin{equation}
\left\{
\begin{array}{rcl}
%\label{deltaXquinquies}
\dot{R}^\mu
&=&
\displaystyle
P^\mu-\frac{1}{(g+1)}\frac{1}{S_{\alpha\beta}F^{\alpha\beta}}\Big[(g-2)\,{S}^{\mu\nu}F_{\nu\rho}{}P^\rho
-g\,{S}^{\mu\nu}\p_{\nu}F_{\rho\sigma}\,{S}^{\rho\sigma}
\Big],
\\[12pt]
%\label{deltaPquinquies}
\dot{P}^\mu
&=&
\displaystyle
-eF^\mu_{\;\nu}\,\dot{R}^\nu-\frac{eg}{4}\,\p^{\mu}F_{\rho\sigma}\,{S}^{\rho\sigma},
\\[12pt]
%\label{deltaSquinquies}
\dot{S}^{\mu\nu}
&=&
\displaystyle
P^\mu\dot{R}^\nu-P^\nu\dot{R}^\mu+\frac{eg}{2}\,\Big[{S}^{\mu}_{\;\rho}\,F^{\rho\nu}-{S}^{\nu}_{\;\rho}\,F^{\rho\mu}\Big].
\end{array}
\right.
\label{Qgeqmot}
\end{equation}

\goodbreak

These equations, which reduce to (\ref{mincoupsigma})
for $g=0$, constitute the zero-rest-mass counterparts of the celebrated Bargmann-Michel-Telegdi equations for  massive relativistic particles \cite{BMT}, as well as $4$ dimensional analogs of ``exotic'' anyons in the plane \cite{AnAn}. In the normal case, $g=2$, resulting from the Dirac equation \cite{Duval75}, 
the previously considered anomalous velocity is canceled but there arises a new, Stern - Gerlach-type contribution involving the derivative of the external electromagnetic field. Thus, an anomalous velocity term shows up for any value of the gyromagnetic ratio $g$.
 
Now we turn to a $(3+1)$-decomposition. Things behave as before, up to some subtle differences.
Firstly, we have
\beq
R=(\br, t),\qquad
P=(\bp, \cE),\qquad
S_{j4}=\left(\frac{\bp}{\cE}\times\bs\right)_j,
\label{new31}
\eeq
where the spin tensor is still defined as in (\ref{spintensor}), but
the new dispersion  relation generalizes (\ref{Table}) \footnote{As the term $S\cdot{}F$ itself involves $\cE$, Eq. (\ref{newdispers}) is a third-order algebraic equation for $\cE$.}, namely
\beq
\cE=\sqrt{|\bp|^2-\frac{eg}{2}S\cdot{}F\,
}\, .
\label{newdispers}
\eeq 

Decomposing the electromagnetic field into its electric and magnetic components, the quantity (\ref{regconds}) (a) is generalized to
\beq
\half S\cdot{}F = 
 \bs\cdot\left(\bB-\frac{\bp}{\cE}\times\bE\right).
\eeq

\goodbreak

Then a rather tedious calculation yields the following $(3+1)$-form of the equations of motion (\ref{Qgeqmot}), namely
\begin{equation}
\left\{
\begin{array}{rcl}
\dot{\br}
&=&
 \displaystyle\frac{3g}{2(g+1)}\,\bp-\left(\frac{g-2}{g+1}\right)\left[\displaystyle\frac{\bs\cdot\bp}{S\cdot{}F}
 \left(\bB-\frac{\bp}{\cE}\times\bE\right)-
 \frac{eg}{2}\bE\times\frac{\bs}{\cE}\right]
\\[12pt]
&&-\,\displaystyle\frac{g}{2(g+1)S\cdot{}F}\,
\Big(\bs\times(S\cdot\bD{F})-\frac{\bp}{\cE}\times\bs\,(S\cdot{}D_tF)\Big),
\\[20pt]
\dot{t}
&=&
\displaystyle\frac{g}{2(g+1)\cE}\big(3|\bp|^2-(g+1)eS\cdot{}F\big)
-\left(\frac{g-2}{g+1}\right)\frac1{{\cE}\,S\cdot{}F}\, ({\bp\cdot\bB})(\bs\cdot\bp)
\\[12pt]
&&+\displaystyle\frac{eg(g-2)}{2(g+1){\cE}^2}\,\bs\cdot(\bp\times\bE)
-\displaystyle\frac{g}{(g+1)}
\frac1{\cE\,S\cdot{}F}
(\bp\times{\bs})\cdot(S\cdot\bD{F}),
 \\[20pt] 
\dot{\bp}
&=&e\big(\bE\,\dot{t}+\dot{\br}
\times\bB\big)+\displaystyle\frac{eg}{4}S\cdot\bD{F},
\\[20pt]
\dot{\bs}
&=& \displaystyle\bp\times\dot{\br}
+
\displaystyle\frac{eg}{2}
\left(\left(\frac{\bp}{\cE}\times\bs\right)\times\bE+\bs\times\bB\right),
\\
\end{array}
\right.
\label{tgeqmot}
\end{equation} 
where we introduced the new shorthands
\beq
S\cdot{}D_{j}F=2\bs\cdot\left(\p_j\bB-\frac{\bp}{\cE}\times\p_j\bE\right),
\qquad
S\cdot{}D_tF=2\bs\cdot\left(\p_t\bB-\frac{\bp}{\cE}\times\p_t\bE\right).
\label{Dfields}
\eeq

\goodbreak

When $g=0$ we recover (\ref{tg0eqmot}). 

To gain more insight, we consider the case $g=2$  and assume that the external fields are constant \footnote{Eqns. (\ref{Qgeqmot}) imply that when the electromagnetic field is a constant, $S.F$ in the denominator is a constant of the motion. The system is therefore regular whenever the initial conditions are regular.}. Then the field-derivative terms drop out as does also the anomalous velocity term \footnote{This is also what happens in the plane for  for $g=2$, just like \cite{AnAn}.},
and the complicated system (\ref{tgeqmot}) simplifies to one reminiscent of a massive relativistic particle, 
\beq
(g=2)\qquad
\left\{\barraynb{cll}
\cE\, 
 \displaystyle\frac{d\br}{dt}&=&\displaystyle{\bp},
\\[12pt]
\displaystyle\frac{d\bp}{dt}&=&e\left(\bE+\displaystyle\frac{\bp}{\cE}\times\bB\right),
\\[12pt]
\displaystyle\frac{d\bs}{dt}&=&\displaystyle\frac{e}{\cE}
\left(\big(\displaystyle\frac{\bp}{\cE}\times\bs\big)\times\bE
+\displaystyle{\bs}\times\bB\right),
\earraynb\right.
\label{cfg2}
\eeq 
assuming that $\cE\neq0$, which acts as a sort of effective mass, is real.
(Recall that $\bp\neq0$ implies that $\cE$ can not vanish).

In a \emph{pure magnetic field} momentum and spin 
 satisfy equations of identical form,
\beq
\frac{d\bp}{dt}=\frac{e}{\cE}\,\bp\times\bB,
\qquad
\frac{d\bs}{dt}=\frac{e}{\cE}\,\bs\times\bB.
\label{g2pureBeq}
\eeq
Thus
\beq
\left\{\barraynb{cc}
|\bp|=\const\neq0, &\bp\cdot\bB=\const,
\\[4pt]
|\bs|=\const\neq0, &\bs\cdot\bB=\const,
\earraynb\right.
\quad
\implies
\quad
\left\{\barraynb{cll}
p_z&=&\const,\qquad s_z=\const,
\\[4pt]
\cE&=&\sqrt{|\bp|^2-e\bs\cdot\bB}=\const.
\earraynb\right.
\eeq
Choosing the $z$ axis in the direction of the magnetic field, $\bB=B\hat{\bz}$,
for example, the momentum and spin vectors precess  
and the position spirals around the $z$ axis
with common angular velocity $\omega=-eB/\cE$,
\beq
\bp(t)=(p_0e^{-i(eB/\cE)t},p_z),
\;
\bs(t)=(s_0e^{-i(eB/\cE)t},s_z),
\;
\br(t)=\Big(\frac{ip_0}{eB}\,e^{-i(eB/\cE)t},\,\frac{p_z}{\cE}t\Big)+\br_0,\;
\label{g2pureB}
\eeq
where $p_0=p_x+ip_y,\,s_0=s_x+is_y$, 
cf. Fig. \ref{g2plot}. It is worth noting that for weak fields and pure magnetic field, and $\bs=\half\hbp$ ,
\beq
\cE\approx |\bp|-\frac{eg}{4}S\cdot{}F
=|\bp|-e\,\frac{\hbp\cdot\bB}{2|\bp|},
\label{weakfielddisp}
\eeq
which is the modified dispersion relation proposed  in \cite{SonYama2,ChenSon,Manuel}.\vskip-3mm
%%%%%%%%%
\begin{figure}[h]
\begin{center}
\includegraphics[scale=.44]{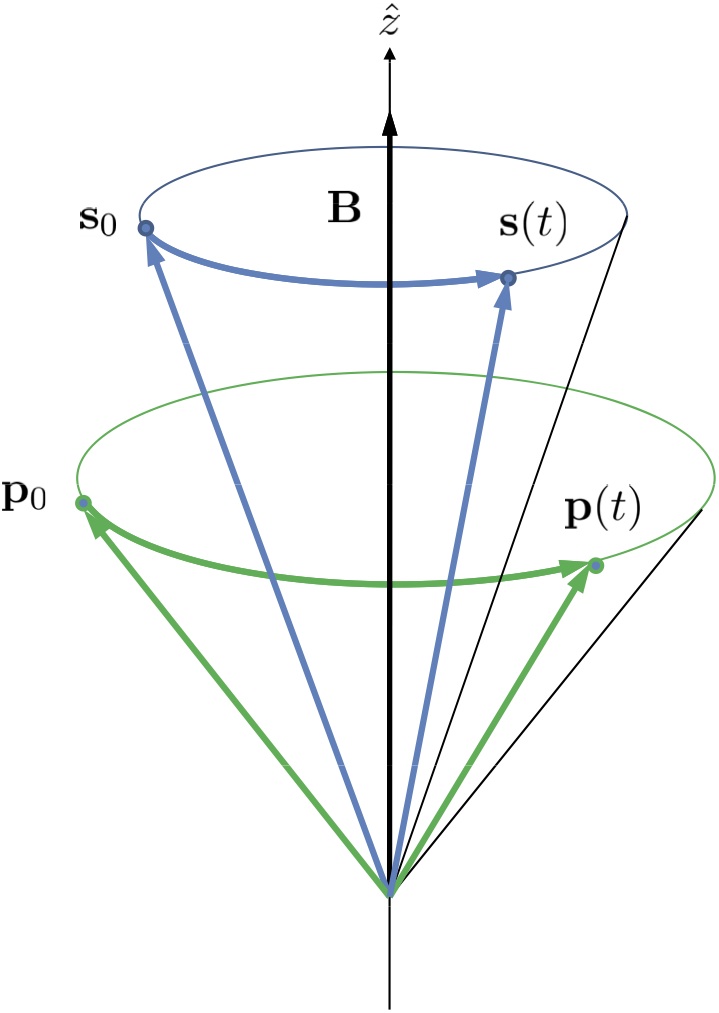}
\qquad
\includegraphics[scale=.57]{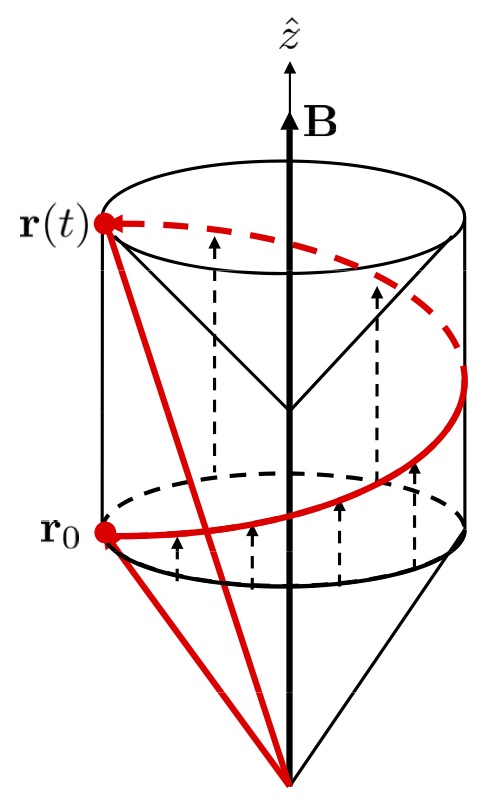}
\\
\hskip13mm
(a)\hskip57mm (b)
\vspace{-8mm}
\end{center}
\caption{\it (a) Motion in pure constant  magnetic field $\bB$. Both  momentum $\green{\bp}(t)$ and spin $\blue{\bs}(t)$ precess, with identical angular velocity, around the $\bB$ direction,  cf. eqn (\ref{g2pureB}). (b) The position $\red{\br}(t)$ spirals on a cylinder around the $\bB$-axis, obtained by combining precession with the vertical drift of the supporting cone itself.}
\label{g2plot}
\end{figure}
%%%%%%%%%
 
By eqns (\ref{g2pureB})
the direction of the rotation is reversed if the sign of the electric charge $e$ is reversed, whereas the direction of the vertical propagation is unchanged. 
As to chirality, when the sign of  spin is reversed, eqns (\ref{g2pureB}) 
  yield similar spiraling motions, rotating and drifting
in the same directions but with different angular velocities,
namely with
\beq
\omega_{\pm}=-\frac{eB}{\cE_{\pm}}=-\frac{eB}{\sqrt{|\bp|^2\mp{e}\hbp\cdot\bB}}
\approx
-\frac{eB}{|\bp|}\left(1\pm\frac{e\hbp\cdot\bB}{2|\bp|}\right),
\label{omegapm}
\eeq 
assuming that the magnetic field is weak.  

In the purely magnetic case, enslavement (\ref{enslavement})  \emph{can} be consistently  required. However,
this is \emph{manifestly not so} in the presence of an electric field \footnote{This behavior is once again related to \emph{gauge covariance}: a constant electric field could be eliminated by a boost --- but
boost freedom and enslavement seem to be incompatible.}:  \emph{the independent spin degree of freedom can not be switched off} if $\bE\neq0$.

%%%%%%%%%%%%%%%%%%%%%%%%%%%%%%%%%%%%%%%%%%%%%%%%%%%%%%%%%%%%%%%%%%%%%%%%%%%%%%%%%%%%%%%%%%%%%%%%%
\section{Conclusion}\label{ConcSect}
%%%%%%%%%%%%%%%%%%%%%%%%%%%%%%%%%%%%%%%%%%%%%%%%%%%%%%%%%%%%%%%%%%%%%%%%%%%%%%%%%%%%%%%%%%%%%%%%%

In this paper we have shown that the semiclassical chiral fermion model, much discussed in recent times in connection with the chiral magnetic and chiral vortical effects \cite{SonYama1,hadrons,SonYama2,Stephanov,ChenWang,Dunne,Stone,Chen2013,ChenSon}, can, in the free case,  be related to the zero mass and spin-$1/2$ particle model of \cite{SSD}. The latter carries a natural Poincar\'e symmetry that  can be exported to  chiral  fermions using  the above  relation.

 We obtain, in particular, Lorentz boosts as proposed  recently    \cite{ChenSon}. 
One could argue that this is what one would expect for a relativistic theory; we would like to stress, however, that this action is \emph{not} the usual, natural one on ordinary 
space-time --- on the contrary, it resembles a dynamical symmetry in that it also involves the momentum. We contend that  the variable $\bx$,  viewed commonly as position,  does not transform correctly under a boost; it is rather our $\br$, which is
the \textit{bona fide} position coordinate studied in Sections \ref{masslessSec} and \ref{chirsymm}. The situation is reminiscent of that of Newton-Wigner coordinates, familiar for the Dirac equation.

Our model is \emph{similar to but different from} those proposed in \cite{SonYama1,hadrons,Stephanov,SonYama2,ChenWang,Dunne, Stone,Chen2013,ChenSon,Manuel}~:
while the usual chiral model (\ref{chiract}) has no independent spin variable,
 ours has  additional degrees of freedom associated with unchained spin and instrumental for having 
 a natural Poincar\'e action. These additional degrees of freedom do not influence the free dynamics,  though, as
they can be eliminated by enslaving the spin to the momentum, using the additional symmetry referred to as Wigner (-Souriau) translations \cite{Wigner,Penrose67,SSD,NewStone,WS}.

The  models become even more different when coupled to an external field~:
the standard chiral models have a $6$-dimensional phase space, whereas ours has, in the coupled case, $8$ dimensions. Also the motions appear rather different in the two frameworks. The difference comes from choosing the  physically relevant position coordinate: $\bx$ in the chiral model and~$\br$ in the one we propose here. The  question is not purely academic, since the coupling to a field is expressed precisely in terms of the position. 

The difference between the theories originates in that in the usual approach \cite{SonYama1,Stephanov,Stone,Dunne, SonYama2,hadrons, ChenWang,Chen2013,ChenSon,Manuel,NewStone,WS}  they  are  derived from some  
widely accepted and physically trusted theory like the Dirac equation, transport theory, fluid dynamics, etc, while we build ours from the principles of Souriau's mechanics, based on group theory, cf. \cite{SSD,Souriau74}. 
  
We present our investigations in symplectic, instead of usual variational terms. Although the two frameworks are essentially equivalent \cite{SSD,VarSpin}, the symplectic one 
 is better adapted to study degenerate systems as  in the free case.
 An alternative point of view is presented in \cite{WS}. 

Non-Abelian generalization could also be considered along the lines discussed in \cite{DuvalAix}.

%%%%%%%%%%%%%%%%%%%%%%%
\begin{acknowledgments} 
We would like to thank Mike Stone for calling our attention to this problem, for enlightening correspondence and for sending us his papers \cite{NewStone}.
``Information tunneling'' may have mutually influenced our work.
Pengming Zhang helped us to produce our figures.
We are moreover indebted to Fran\c{c}ois Ziegler, Gary Gibbons, Kostya Bliokh and L\'aszl\'o Feh\'er for stimulating discussions and correspondence. 
\end{acknowledgments}

\goodbreak

%%%%%%%%%%%%%%%%%%%%%%%%%%%%%%%%%%%%%%%%%%%%%%%%%%%%%%%%%%%%%%%%%%%%%%%%%%%%%%
%%%%%%%%%%%%%%%%%%%%%%%%%%%%%%%%%%%%%%%%%%%%%%%%%%%%%%%%%%%%%%%%%%%%%%%%%%%%%%

\renewcommand\thesection{A}
\numberwithin{equation}{section}

%%%%%%%%%%%%%%%%%%%%%%%%%%%%%%%%%%%%%%%%%%%%%%%%%%%%%%%%%%%%%%%%%%%%%%%%%%%%%%
%%%%%%%%%%%%%%%%%%%%%%%%%%%%%%%%%%%%%%%%%%%%%%%%%%%%%%%%%%%%%%%%%%%%%%%%%%%%%%
\section{Appendix : Souriau's  mechanics as generalized variational calculus
}\label{AppendixA}
%%%%%%%%%%%%%%%%%%%%%%%%%%%%%%%%%%%%%%%%%%%%%%%%%%%%%%%%%%%%%%%%%%%%%%%%%%%%%%
%%%%%%%%%%%%%%%%%%%%%%%%%%%%%%%%%%%%%%%%%%%%%%%%%%%%%%%%%%%%%%%%%%%%%%%%%%%%%%

In  the framework of \cite{SSD}  the dynamics is determined by a
closed two-form $\sigma$ of constant rank defined on some evolution space $V$;  
the motions, described by curves or by surfaces of $V$, are the so-called ``characteristic leaves'', tangent to the kernel, $\ker\sigma$, of the two-form~$\sigma$. 

To explain how this comes about, we consider a particle described by a Lagrangian on phase space of which (\ref{chiract}) is an example that will serve as an illustration. Denoting the phase space variables $\bx$ and $\bp$ collectively by $\xi=(\xi^\alpha)$,  the Lagrangian in (\ref{chiract}) is of the form $u_\alpha\dot{\xi}^\alpha-h(\xi)$ and  the associated variational equations are 
\beq
\omega_{\alpha\beta}\dot{\xi}^\beta=\p_\alpha{h},
\qquad\hbox{where}\qquad
\omega_{\alpha\beta}=\p_\alpha{u}_\beta-\p_\beta{u}_\alpha.
\label{phspvareq}
\eeq
We note \textit{en passant} that if  the matrix $(\omega_{\alpha\beta})$ is regular, then multiplying (\ref{phspvareq})
with the inverse matrix would yield Hamilton's equations.

A  next step is to extend the $6$-dimensional phase space into 
the $7$-dimensional \emph{evo\-lution space} $V^7$ described by triples $y=(\bx,\bp,t)$ and unify  the two-form $\omega=\half\omega_{\alpha\beta} d\xi^\alpha\wedge d\xi^\beta$ with the Hamiltonian into the two-form
\beq
\sigma=\omega-dh\wedge dt.
\label{Souriauform}
\eeq
Then the equations of motion (\ref{phspvareq}) become finally
\beq
\sigma(\dot{y},\,\cdot\,)=0,
\label{kersigma}
\eeq
expressing that the velocity, $\dot{y}$, of the motion unfolded into the evolution space  belongs to $\ker{\sigma}$, see Fig. \ref{EvolSpace}.
%%%%%%%%%
\begin{figure}[hpt]
\begin{center}\vspace{-3mm}
\includegraphics[scale=.55]{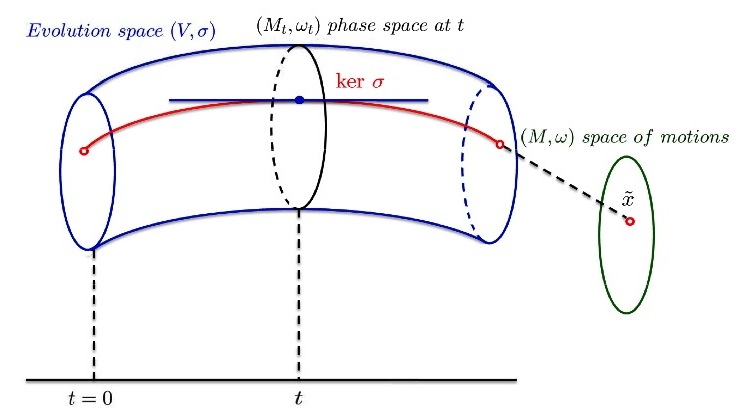}
\vspace{-10mm}
\end{center}
\caption{\it A motion is described,  in the evolution space $V$, by a submanifold, which is tangent to the kernel of a closed two-form, $\sigma$, of constant rank. The space of motions is the symplectic manifold, $(M,\omega)$, obtained from $(V,\sigma)$  by factoring out these  characteristic submanifolds.
A phase space at time $t$ is a section of the
evolution space for a fixed value of $t$, and provides a
local chart of the space of motions.
} 
\label{EvolSpace}
\end{figure}
%%%%%%%%%

More generally, we can consider an evolution space $V$ of dimension $d$, endowed with a closed two-form $\sigma$ of constant rank; let the kernel of $\sigma$ be an $r$-dimensional vector space.
Then general theorems guarantee that $\ker{\sigma}$ is tangent, at each point, to an $r$-dimensional submanifold called a \emph{characteristic leaf} of $\sigma$;
the latter can be viewed as \emph{solution of 
a generalized variational problem}. 

Factoring out the characteristic leaves provides us with
a symplectic manifold $(M,\omega)$ of dimension $2n=d-r$, called the \emph{space of motions}, which can be regarded as an abstract substitute  for the phase space. For further details the reader is invited to consult, e.g.,~\cite{SSD,VarSpin}.

We just mention that a \emph{symmetry} is a transformation of the evolution space $V$ which preserves
its two-form $\sigma$. The relation between symmetries and conservation laws is established by the symplectic form of Noether's theorem. 
Conversely, the space of motions of classical systems with a given symmetry can be constructed, under suitable conditions, by group theoretical considerations \cite{SSD}. 

\renewcommand\thesection{B}
\numberwithin{equation}{section}

%%%%%%%%%%%%%%%%%%%%%%%%%%%%%%%%%%%%%%%%%%%%%%%%%%%%%%%%%%%%%%%%%%%%%%%%%%%%%%
%%%%%%%%%%%%%%%%%%%%%%%%%%%%%%%%%%%%%%%%%%%%%%%%%%%%%%%%%%%%%%%%%%%%%%%%%%%%%%
\section{Appendix : Massless, spinning relativistic particle models
}\label{AppendixB}
%%%%%%%%%%%%%%%%%%%%%%%%%%%%%%%%%%%%%%%%%%%%%%%%%%%%%%%%%%%%%%%%%%%%%%%%%%%%%%
%%%%%%%%%%%%%%%%%%%%%%%%%%%%%%%%%%%%%%%%%%%%%%%%%%%%%%%%%%%%%%%%%%%%%%%%%%%%%%

We recall, here, the model of massless, spinning particle dwelling in Minkowski space-time, as spelled out in \cite{SSD}, Section (14.29).
The space of motions (or of classical states) of a free relativistic particle is a homo\-geneous symplectic manifold of the Poincar\'e group, $\rE(3,1)$. 
These coadjoint orbits are known and classified; those with zero mass, and nonzero spin are constructed as follows.

For convenience we deal with the neutral Poincar\'e group $G=\SE_+(3,1)$ whose elements are pairs $g=(\cL,\cC)$ with $\cL\in\SO_+(3,1)$, and $\cC\in\bbR^{3,1}$, a space-time translation. Every element of the dual $\se(3,1)^*$ of the Lie algebra, $\se(3,1)$, of $G$ is a pair $\mu=(M,P)$ with $M\in\sort(3,1)$, the Lorentz momentum, and $P\in\bbR^{3,1}$, the linear momentum. The pairing between these spaces is given by $\mu\cdot{}Z=\half{}M_{\mu\nu}\Lambda^{\mu\nu}-P_\mu\Gamma^\mu$ with $Z=(\Lambda,\Gamma)\in\se(3,1)$.

We will deal with oriented and time-oriented Lorentz frames $E=(I,J,K,L)$ of Minkowski space-time such that the only nonzero scalar products are $I_\mu{}J^\mu=K_\mu{}K^\mu=L_\mu{}L^\mu=-1$, with $I$ (null) future-pointing. It is useful to identify those frames with the neutral Lorentz group via $E=\cL{}E_0$, where $E_0$ is some fixed frame, as well as space-time translations, $\cC$, with Minkowskian events, $R$.

Picking then a fixed Poincar\'e-momentum $\mu_0=(M_0,P_0)$ such that $M_0=s I_0\times{}J_0$ (the cross-product of $I_0$ and~$J_0$, i.e., $(M_0)_{\mu\nu}=-s\,\epsilon_{\mu\nu\rho\sigma}\,I_0^\rho{}J_0^\sigma
$) with $s>0$ interpreted as the classical spin, and $P_0=I_0$, we may define the one-form 
\begin{equation}
\alpha=\mu_0\cdot g^{-1}dg
\label{alpha}
\end{equation}
on $G$. Then, as a general result, the two-form 
\begin{equation}
\sigma=d\alpha
\label{sigma=dalpha}
\end{equation}
descends to the coadjoint orbit $M=G/G_{\mu_0}$ as its canonical symplectic form; the leaves generated by the stabilizer $G_{\mu_0}$ of $\mu_0$ are interpreted as the motions of our (free) particle and integrate, by construction, the null distribution, $\ker\sigma$, on $(G,\sigma)$. These leaves project down to space-time as the worldsheets of our particle.

In the case under study the one-form (\ref{alpha}) of $G$ reads
\beq
\alpha=-I_\mu\,dR^\mu+s K_\mu\,dL^\mu,
\eeq
whereas its derivative (\ref{sigma=dalpha}) descends to the evolution space $V^9=G/\SO(2)$ 
in (\ref{evolutionspace}), the $\SO(2)$-action on $G$ being $(I,J,K,L,R)\to(I,J,K\cos\theta+L\sin\theta,-K\sin\theta+L\cos\theta,R)$.
This two-form, still denoted by $\sigma$ with a slight abuse of notation, is finally given by (\ref{freespinsigma}) where we have put $P=I$ for the four-momentum, and $S=s I\times{}J$ for the spin tensor. 

\goodbreak

\renewcommand\thesection{C}
\numberwithin{equation}{section}

%%%%%%%%%%%%%%%%%%%%%%%%%%%%%%%%%%%%%%%%%%%%%%%%%%%%%%%%%%%%%%%%%%%%%%%%%%%%%%
%%%%%%%%%%%%%%%%%%%%%%%%%%%%%%%%%%%%%%%%%%%%%%%%%%%%%%%%%%%%%%%%%%%%%%%%%%%%%%
\section{Appendix : Finite coadjoint action of the Poincar\'e group}\label{AppendixC}
%%%%%%%%%%%%%%%%%%%%%%%%%%%%%%%%%%%%%%%%%%%%%%%%%%%%%%%%%%%%%%%%%%%%%%%%%%%%%%
%%%%%%%%%%%%%%%%%%%%%%%%%%%%%%%%%%%%%%%%%%%%%%%%%%%%%%%%%%%%%%%%%%%%%%%%%%%%%%

\renewcommand\thesection{C}
\numberwithin{equation}{section}

We recall that a Lorentz transformation of $\bbR^{3,1}$ is of the form
\begin{equation}
\cL=\exp\left(
\begin{array}{cc}
0&\alpha\bu
\\[4pt]
\alpha\bu^{T}&0
\end{array}
\right)
.
\left(
\begin{array}{cc}
A&0\\[4pt]
0&1
\end{array}
\right)\in\SO_+(3,1),
\label{L}
\end{equation}
where $\alpha\in\bbR$ is the rapidity and $\bu\in{}S^2$ the direction of the boost, $\bb=\tanh\alpha\,\bu$, and $A\in\SO(3)$; 
we  put $\gamma=\cosh\alpha=(1-|\bb|^2)^{-\half}$. 
Using the shorthand  $B=(B^i_j)=(\delta^i_j+(\gamma-1)u^i\,u_j)$, an element of the connected (also called neutral) Poincar\'e group is of the form
\begin{equation}
g=\left(
\begin{array}{ccc}
BA&\gamma\,\bb&\bc
\\[4pt]
\gamma\bb^{T}A&\gamma\,&e
\\[4pt]
0&0&1
\end{array}
\right)\in\SE_+(3,1),
\label{LCbis}
\end{equation}
where $\bc\in\bbR^3$ is a space-translation, and $e\in\bbR$ a time-translation. The Lie algebra of the Poincar\'e group is therefore spanned by the matrices 
\begin{equation}
Z=\left(
\begin{array}{ccc}
\widetilde{\bomega}
&\bbeta&\bgamma\\[6pt]
\bbeta^T&0&\varepsilon\\[6pt]
0&0&0
\end{array}
\right)\in\se(3,1)
\label{Z}
\end{equation}
where $\widetilde{\bomega}\in\sort(3)$ is identified with $\bomega\in\bbR^3$, also $\bbeta,\bgamma\in\bbR^3$ and $\varepsilon\in\bbR$. Then the  adjoint action $Z\to{}Z'=(\bomega',\bbeta',\bgamma',\epsilon')=\mathrm{Ad}(g^{-1})Z$ reads
\begin{eqnarray}
\bomega'&=&A^{T}\big(\gamma\bomega-(\gamma-1)\bu(\bu\cdot\bomega)+\gamma\bb\times{}B\bbeta\big)
\\
\bbeta'&=&A^{T}\big(\gamma{}A(\bomega\times\bb)-\gamma^2\bb(\bb\cdot\bbeta)+\gamma{}A\bbeta\big)
\\ 
\bgamma'&=&A^{T}\big(A(\bomega\times\bc)-\gamma\bb(\bc\cdot\bbeta)+A\bbeta{}\,e+A\bgamma-\gamma\bb\,\varepsilon\big)
\\ 
\varepsilon'&=&\gamma\big((\bb\times\bc)\cdot\bomega+\bbeta\cdot\bc-(\bb\cdot\bbeta){}\,e-\bb\cdot\bgamma+\varepsilon\big).
\end{eqnarray}
Denoting by $\mu=(\belle,\bg,\bp,\cE)$ a ``moment'' in $\se(3,1)^*$ where $\mu\cdot{}Z=\belle\cdot\bomega-\bg\cdot\bbeta+\bp\cdot\bgamma-\cE\,\varepsilon$, we then find the coadjoint representation
$\mu\to\mu'=\mu\circ\mathrm{Ad}(g^{-1})$ where
\begin{eqnarray}
\nonumber
\belle'&=&\gamma{}A\belle-(\gamma-1)(\,\bu{\cdot}A\belle)\bu-\gamma\bb\times{}A\bg+\bc\times{}A\bp\\
\label{ellp}
&&-\gamma{\cE}\,\bb\times\bc{}-(\gamma-1)\,(\bu{\cdot}A\bp)\,\bu\times\bc
\\
\label{gp}
\nonumber
\bg'&=&\gamma\bb\times{}A\belle+\gamma{}A\bg-(\gamma-1)\big(\bu{\cdot}A(\bg+\bp{}e)\big)\,\bu+\gamma(\bb{\cdot}A\bp)\,\bc 
\\
&&-{e}\,A\bp{}+\gamma{}\cE(\bc-\bb{}e)
\\ 
\label{pp}
\bp'&=&A\bp+(\gamma-1)(\bu\cdot{A}\bp)\,\bu+\gamma{}\cE\,\bb
\\ 
\label{Ep}
\cE'&=&\gamma(\cE+ \bb{\cdot}A\bp).
\end{eqnarray}
At last, restricting ourselves to positive  helicity and  energy, the $\SE_+(3,1)$-action is expres\-sed in terms of the quantities $(\tbx,\tbp)$ describing the space of motions $(M^6,\omega)$ given in (\ref{Table}); the Poincar\'e-action (\ref{Ponpx}) follows then at once.

\end{document}